\documentclass[english,twocolumn, notitle, superscriptaddress,amsmath,amssymb,showpacs,showkeys]{revtex4-1}
\usepackage[T1]{fontenc}
\usepackage{amsmath}
\usepackage{amssymb}
\usepackage{graphicx}

\makeatletter

\providecommand{\tabularnewline}{\\}
\newcommand{\prlsec}[1]{\textit{#1.---}}

\makeatother

\usepackage{babel}
\begin{document}

\title{Dynamical model selection near the quantum-classical boundary}

\author{Jason F. Ralph}
\email{jfralph@liverpool.ac.uk}

\selectlanguage{english}%

\affiliation{Department of Electrical Engineering and Electronics, University
of Liverpool, Brownlow Hill, Liverpool, L69 3GJ, UK.}

\author{Marko Toro\v{s}}
\email{m.toros@soton.ac.uk}

\affiliation{Department of Physics and Astronomy, University of Southampton, University
Road, Southampton, SO17 1BJ, UK}

\author{Simon Maskell}
\email{smaskell@liverpool.ac.uk}

\selectlanguage{english}%

\affiliation{Department of Electrical Engineering and Electronics, University
of Liverpool, Brownlow Hill, Liverpool, L69 3GJ, UK.}

\author{Kurt Jacobs}
\email{kurt.jacobs@umb.edu}

\selectlanguage{english}%

\affiliation{U.S. Army Research Laboratory, Computational and Information Sciences
Directorate, Adelphi, Maryland 20783, USA.}

\affiliation{Department of Physics, University of Massachusetts at Boston, Boston,
MA 02125, USA}

\affiliation{Hearne Institute for Theoretical Physics, Louisiana State University,
Baton Rouge, LA 70803, USA}

\author{Muddassar Rashid}
\affiliation{Department of Physics and Astronomy, University of Southampton, University
Road, Southampton, SO17 1BJ, UK}

\author{Ashley J. Setter}

\affiliation{Department of Physics and Astronomy, University of Southampton, University
Road, Southampton, SO17 1BJ, UK}

\author{Hendrik Ulbricht}
\affiliation{Department of Physics and Astronomy, University of Southampton, University
Road, Southampton, SO17 1BJ, UK}

\date{\today}
\begin{abstract}
We discuss a general method of model selection from experimentally recorded time-trace data. This method can be used to distinguish between quantum and classical dynamical models. It can be used in post-selection as well as for real-time analysis, and offers an alternative to statistical tests based on state-reconstruction methods. We examine the conditions that optimize quantum hypothesis testing, maximizing one's ability to discriminate between classical and quantum models. We set upper limits on the temperature and lower limits on the measurement efficiencies required to explore these differences, using a novel experiment in levitated optomechanical systems as an example.
\end{abstract}
\maketitle

\prlsec{Introduction} There are a number of ways in which a system can be determined to be quantum mechanical. Typically, the system must be isolated from extraneous noise and operated at very low temperatures, so that the system is in a ground state or another low lying energy state. The system can be subjected to a series of individual or joint measurements to build up a picture of the state (as in interference experiments and state tomography \cite{Nai2003,Ger2011,Smi1993,Lvo2001,Roo2004,Res2005}) or manipulated using an external field to demonstrate superposition states (such as avoided crossings in the observed energy spectra \cite{Nak1997,Fri2000,Wal2000,Mar2002}). These experiments provide direct evidence of quantum behavior but they can be difficult to perform when the system has several degrees of freedom and large numbers of measurements are required. 

More efficient alternatives have been devised with the growth of quantum information as a subject area. Specific sequences of measurements can be applied to ascertain whether the system contains non-classical correlations (entanglement) associated with quantum behavior \cite{Bar2003,Bou2004,Lu2007}. {\em Entanglement witnesses} do not necessarily allow an experimentalist to quantify the degree of entanglement, but they do allow her to say that entanglement is present and, hence, that the system is quantum mechanical rather than classical. All of these methods are intended to provide direct evidence that the system is manifestly non-classical, e.g. discrete energy levels, interference, superposition states, and entanglement.

An alternative approach is to try to determine whether the system {\em dynamics} are quantum rather classical. An elegant approach to this task is to use the technique of {\em quantum hypothesis testing} \cite{Tsa2012,Tsa2013}. In situations where direct experiments are not possible, or are beyond the reach of current experiments, this method can also be used to inform future work, explore regions of parameter space, and to focus experimental efforts. This is the motivation for the current paper.  

In this letter, we use the quantum hypothesis testing approach, often referred to as {\em model selection} in classical Bayesian inference~\cite{Gor2002}, to construct a general method of model selection, which is an alternative to state-reconstruction based statistical tests. We reformulate the problem as a Neyman-Pearson decision rule and quantify the accuracy of the selected model using the confusion matrix. As an example application, we devise a novel experiment for optically levitated systems~\cite{Ash1970,Man1998,Gan1999,Vul2000}, and we optimise the Hamiltonian parameters to enhance the distinguishability of quantum and classical dynamics. The proposed experiment does not require complicated preparation and measurement protocols, but relies only on the detected photo-current~\cite{Tor2018}. Quantum behavior has not yet been established with such massive systems, and improving the understanding of where and when such evidence might be available is an important open question. We demonstrate that two experimental parameters, the effective temperature and the efficiency of the continuous measurement, are critical to the ability to distinguish between quantum and classical stochastic dynamics in this system. 

\prlsec{Dynamical models} A model is composed of three elements: (i) the description of the system, i.e. the state, (ii) the dynamical law, and (iii) the detection process. In this letter, we consider non-relativistic, single-particle, classical and quantum dynamics with a diffusive, Markovian environment, subject to continuous monitoring of the position of the particle. We denote the state by $S_{c}$, the measured time-trace by $I_{\text{exp}}$, and the measured position by $\tilde{q}$ (either classical or quantum). Note, however, that these assumptions are not essential, but only a matter of convenience of presentation, and could, at least in principle, all be relaxed. In particular, the analysis for general, non-relativistic, diffusive, Markovian models can be carried out in full analogy with the analysis presented in this letter (see Supplementary material S1). 

Under these assumptions, the state $S_{c}$ formally evolves according to~\cite{Bel1999,Wis2010,Jac2014} (in It\^{o} form):
\begin{equation}
dS_{c}=\mathcal{K}S_{c}dt+\gamma\mathcal{D}\left[\frac{\tilde{q}}{\sigma}\right]S_{c}dt+\sqrt{\eta\gamma}\mathcal{H}\left[\frac{\tilde{q}}{\sigma}\right]S_{c}dW,\label{eq:dynamics}
\end{equation}
where $\mathcal{K}\,\cdot\,$, $\mathcal{\mathcal{D}}[\tilde{q}]\,\cdot\,$, and $\mathcal{H}[\tilde{q}]\,\cdot\,$, denote the Hamiltonian, diffusive and detection terms~\cite{Wis1993}, respectively, $W$ is a zero mean Wiener process, and $\sigma$, $\gamma$ denote a characteristic length scale, frequency of the experiment, respectively, and $\eta$ is the efficiency of the measurement, which is defined to be the ratio between the power due to the recorded measurement signal relative to other sources of noise. Inefficient measurements may arise from loss of signal or corruption of the signal by additional, unprobed environmental degrees of freedom. The detected signal $I_{\text{exp}}(t)$ during an interval, $t\rightarrow t+dt$, is related to the Wiener process by:
\begin{equation}
I_{\text{exp}}(t+dt)-I_{\text{exp}}(t)=dI_{\text{exp}(t)}=\sqrt{\eta}\mathbb{E}[q]dt+\frac{dW}{\sqrt{\gamma}}\label{eq:results}
\end{equation}
where $\mathbb{E}[\,\cdot\,]$ denotes the expectation value with respect to the state $S_{c}$. 

For a given experimental signal, $I_{\text{exp}}(t)$, the conditional evolution of the state can be found by inverting Eq.~(\ref{eq:results}) to find a series of stochastic increments, $d\tilde{W}(t | dI_{\text{exp}(t)})$, to insert back into Eq.~(\ref{eq:dynamics}). The resultant conditional state, $S_{c}$, describes the knowledge about the state of the system, as derived from the measurement record. In classical state estimation, the stochastic increments are often called the {\em innovation} terms~\cite{Aru2002} because they represent the difference between the actual measurement taken and the measurement expected from the conditioned state during each time increment. 
\begin{table}
\begin{tabular}{|c|c|c|}
\hline 
Symbol & Classical & Quantum\tabularnewline
\hline 
\hline 
$S_{c}$ & $P_{c}(q,p;t)$ & $\hat{\rho}_{c}(t)$\tabularnewline
\hline 
$\mathcal{K}\,\cdot\,$ & $\{H,\,\cdot\,\}_{\text{Pb}}$ & $-\frac{i}{\hbar}[\hat{H},\,\cdot\,]$\tabularnewline
\hline 
$\mathcal{D}[\tilde{q}]\,\cdot\,$ & $-\frac{\hbar^{2}}{8}\frac{\partial^{2}}{\partial p^{2}}\,\cdot\,$ & $\frac{1}{8}[\hat{q},[\hat{q},\,\cdot\,]]$\tabularnewline
\hline 
$\mathcal{H}[\tilde{q}]\,\cdot\,$ & $q-\mathbb{E}_{t}[q]$ & $\left(\hat{q}-\mathbb{E}_{t}[\hat{q}]\right)\,\cdot\,+\hspace{2pt}\text{H.c.}$\tabularnewline
\hline 
$\mathbb{E}_{t}[\,\cdot\,]$ & $\int\int dqdp\,\cdot\,P_{c}(q,p;t)$  & $\text{tr}[\,\cdot\,\hat{\rho}_{c}(t)]$\tabularnewline
\hline 
\end{tabular}\centering
\caption{Quantum and classical dynamical models given by Eqs.~(\ref{eq:dynamics}) and (\ref{eq:results}). $P_{c}$ and $\hat{\rho}_{c}$ denote the conditional probability density and conditional statistical operator, respectively. $q$ and $p$ are the classical phase space variables, i.e. position and momentum, respectively, $q$ ($\hat{q}$) denotes the classical (quantum) position observable, $H$ ($\hat{H})$ denotes the classical (quantum) Hamiltonian observable, and $t$ is the time variable. $\mathbb{E}_{t}[\,\cdot\,]$ denotes the expectation value with respect to the conditional state $S_{c}(t)$ at time $t$. }
\label{table1}
\end{table}

The explicit expression for the terms in Eqs.~(\ref{eq:dynamics}) and (\ref{eq:results}) are given in Table~\ref{table1}. The quantum and the corresponding classical models are related by the following
formal prescription: replace the quantum observables $\hat{O}$ with the corresponding classical observables $O$, and commutators with Poisson brackets, i.e. $[\,\cdot\,,\,\cdot\,]\rightarrow i\hbar\{\,\cdot\,,\,\cdot\,\}_{\text{Pb}}$. Note however that, as far as the model selection is concerned, the classical and quantum models could be very different or even completely unrelated.

\prlsec{Decision rule} We now consider a collection $M$ of dynamical models, which we denote by $m_{k}\in M$ ($k=1,..,N$): these can be either quantum or classical (see Eqs.~(\ref{eq:dynamics}), (\ref{eq:results}) and Table \ref{table1}). Before data collection, we suppose that each model is equally likely, which mathematically translates to setting the a priori probabilities to be equal, i.e. the initial probability of model $m_j$ is $p_0(m_{j})=\frac{1}{M}$. After data collection, the goal is to select the model $m_{j}\in M$ that gives the best description of the collected data, i.e. that fits best the recorded time-trace signal $I_{\text{exp}}$. 

According to the Bayes decision rule, a model $m_{j}$ is the best considered model given the detected signal $I_{\text{exp.}}$, if $\forall k\neq j$ we have $p(m_{j}\vert I_{\text{exp}})>p(m_{k}\vert I_{\text{exp}})$. 
However, in some situations, the data $I_{\text{exp}}$ is insufficient to select a given model with certainty, e.g. two models might have experimental predictions that are not distinguishable. It is then useful to introduce an acceptance region $A  =\{I\vert1-\max [p(m_{k}\vert I)]>\tau\}$, where $I$ denotes all possible signals, $\tau$ is a threshold parameter, and $\text{max}$ denotes the maximisation over $m_{k}\in M$. In the case $I_{\text{exp}}\in A$, we apply Bayes decision rule for minimum error, otherwise we conclude the data is inconclusive, i.e. $I_{\text{exp}}$ is in the so-called rejection region. 

The Bayes decision rule and the acceptance region form a two-stage selection: we can combine these two stages by considering an alternative decision rule. Specifically, we consider the Neyman-Pearson decision rule, which has a built-in acceptance threshold parameter $\mu$ for the likelihood ratio~\cite{Tsa2012,Tsa2013}. Specifically, one selects model $m_{j}$, given the detected signal $I_{\text{exp.}}$, if $\forall k\neq j$ we have:
\begin{equation}
\frac{p(m_{j}\vert I_{\text{exp}})}{p(m_{k}\vert I_{\text{exp}})}>\mu.\label{eq:NPdecision}
\end{equation}
Note that the two decision rules coincide  for $\mu=1$, $\tau=1$, and under the assumption of equal a priori probabilities. In the rest of this letter, we will use the latter, more compact rule, given by Eq.~(\ref{eq:NPdecision}).

To apply the decision rule, we are left to specify how to obtain $p(m_{k}\vert I_{\text{exp}})$. Without loss of generality, we assume that the time-trace is given from $t=0$ to $t=t'$, and that the detector has a finite integration time $\Delta t$, such that $n\Delta t=t'$. We now use the property of pairwise independence of the detected signals in each interval $\Delta t$ to obtain an update equation for each of the model probabilities~\cite{Gor2002}:
\begin{equation} \label{eq:update}
p(m_{k}\vert I_{\text{exp},0:t'})\propto p_0(m_{k})\prod_{t = \Delta t:t'} p(\Delta I_{\text{exp},t}\vert m_{k},I_{\text{exp},0:t-\Delta t}) 
\end{equation}
where $I_{\text{exp},0:t'}$ on the left hand-side denotes the total time trace from time $t=0$ to $t=t'$, $p_0(m_{k})$ is the initial probability assigned to the model $m_{k}$, and $\Delta I_{\text{exp},t}$ on the right hand-side is the signal in the interval $[t,t-\Delta t]$. The probability updates are generated using:
\begin{equation}
p(\Delta I_{\text{exp},t}\vert m_{k}, I_{\text{exp},0:t-\Delta t})=\frac{1}{\sqrt{2\pi\Delta t}}\text{exp}\left(-\frac{(\Delta W_{t}^{(m_k)})^{2}}{2\Delta t}\right) \label{eq:gaussian}
\end{equation}
where the increments (innovations) are given by:
\begin{equation}
\Delta W_{t}^{(m_k)}=\sqrt{\gamma}\left(\Delta I_{\text{exp},t}-\sqrt{\eta}\mathbb{E}_{t}[q \vert m_k]\Delta t\right),\label{eq:DeltaW}
\end{equation}
and where $\mathbb{E}_{t}[q \vert m_k]$ is the expected value of $q$, given the dynamical model $m_k$ and the associated conditional state. The probabilities for each of the possible models are updated after each time step using (\ref{eq:update}) and then normalised such that $\sum_k p(m_{k}\vert I_{\text{exp}})=1$. As such, the probabilities being calculated are the relative probabilities between the different dynamical models, which does not necessarily include the possibility of systematic errors. These limitations are considered in detail by Tsang in \cite{Tsa2013}, where the different types of systematic errors are listed and discussed. This limitation does not invalidate the approach presented here. However, it does mean that experimental studies need to be careful to calibrate their systems fully and to verify that systematic errors are either not present, or are included explicitly in one of the dynamical models.

To summarize, given an experimental measurement record consisting of discrete increments $\Delta I_{\text{exp},t}$, and a set of dynamical models $m_k$ describing the possible evolution of the underlying system, we proceed as follows. At $t=0$, set initial probabilities for all models, with the default assumption being that all models are equally likely. At each subsequent time step, 
\begin{enumerate}
\item Calculate the increment $\Delta W_{t}^{(m_k)}$ for each model using (\ref{eq:DeltaW}) and $\Delta I_{\text{exp},t}$, and update the corresponding conditional states using the appropriate form of (\ref{eq:dynamics}).
\item Calculate the probability update $p(\Delta I_{\text{exp},t}\vert m_{k})$ for each model using (\ref{eq:gaussian}).
\item Update all probabilities, using (\ref{eq:update}).
\item Normalize to find relative probabilities. 
\item Repeat using next measurement increment, $\Delta I_{\text{exp},t+\Delta t}$.
\end{enumerate}
Once the updates have been included from all measurements in the record, the decision processing given by (\ref{eq:NPdecision}) can be applied. One benefit of this procedure is that it is clearly iterative, and can therefore be used as an online process, with probabilities being updated as each measurement is taken; or, if required, as a post-processing step after experimental data collection. 

\prlsec{Quality of the decision} We have now introduced dynamical models and selection rules. In particular, we have discussed how to select the best model $m_{j}\in M$ given a measured signal $I_{\text{exp}}$. However,
the model selected might not be overall the best model to describe the experiment, e.g. taking a longer time-trace $I_{\text{exp}}$, or repeating the experiment several times, one might find out that the best model to describe the system is actually a different one. To estimate the probabilities of making a correct or a false selection one can proceed using the following procedure.

Suppose the system evolves according to the model $m_{s}$. In the absence of an experimental record, one can generate a time trace $I_{\text{sim}}^{(m_{s})}$ numerically, solving Eqs.~(\ref{eq:dynamics}) and (\ref{eq:results}), and using a Gaussian random number generator for the Wiener increments $dW$. After the time-trace is generated, one uses the simulated increments to calculate the conditional state evolution for each of the models and generating relative probabilities given the simulated record, $P(m_{k}\vert I_{\text{sim}}^{(m_{s})})$. The most probable model is then selected using the Neyman-Pearson rule given in Eq.~(\ref{eq:NPdecision}), given the simulated time-trace $I_{\text{sim}}^{(m_{s})}$. One repeats this procedure $N^{(s)}$ times to estimate the probabilities of false and correct identification: 
\begin{equation}
P(m_{k}\vert m_{s})\approx\frac{N_{k}^{(s)}}{N^{(s)}},
\end{equation}
where $N_{k}^{(s)}$ denotes the number of times the model $m_{k}$ was selected, when the time-trace $I_{\text{sim}}^{(m_{s})}$ was generated using model $m_{s}$. In the limit $N^{(s)}\rightarrow\infty$, we obtain the probability $P(m_{k}\vert m_{s})$ of selecting model $m_{k}$, when the time trace $I_{\text{sim}}^{(m_{s})}$ has been generated using model $m_{s}$. 

The probabilities $P(m_{k}\vert m_{s})$ form the elements of the so-called confusion matrix $\left(M_{c}\right)_{sk}$~\cite{Han2001}. For example, in the case where we are considering only two models, e.g. a classical one and a quantum one, denoted by $C$ and $Q$ respectively, we can arrange the probabilities of correct and false identification in the following $2\times2$ matrix:
\begin{equation}
M_{c}=\left(\begin{array}{cc}
p({\rm C}|{\rm C}) & p({\rm C}|{\rm Q})\\
p({\rm Q}|{\rm C}) & p({\rm Q}|{\rm Q})
\end{array}\right)
\end{equation}
where $p({\rm C}|{\rm Q})$ is the probability of a Type II error (false negative, assuming that the classical hypothesis ${\rm C}$ is the default or null hypothesis) and $p({\rm Q}|{\rm C})$ is the probability of a Type I error (false positive). More generally, one can generate a Receiver-Operator Characteristic (ROC) curve~\cite{Han2001} by varying the threshold value $\mu$.

\prlsec{Application to optomechanics} Levitated optomechanical systems are a topical area of research. They have been used for ultra-sensitive force measurements \cite{Ger2010}, fundamental tests of gravity \cite{Arv2013}, as well as testing the limits of quantum mechanics \cite{Bas2003,Bas2013,Ber2015}. Here we propose a novel type of experiment to detect non-classical features in such systems using dynamical model selection.

For the purposes of this paper, we assume that the motion of a levitated nanoparticle is decoupled along the three motional axes and discuss only one-dimensional motion~\cite{Tor2018, Mud2016, Mud2017}. Specifically, we suppose that the potential is nonlinear and it forms a Duffing oscillator \cite{Sch1995,Bru1996,Ral2017a}. The Hamiltonian is given by, 
\begin{equation}
\hat{H}=\frac{1}{2}\hat{p}^{2}-\frac{1}{2}\omega^{2}\hat{q}^{2}+\frac{1}{4}\beta\hat{q}^{4}+g\cos(t)\hat{q}\label{DuffHam}
\end{equation}
where we have taken $\hbar=1$, the mass is scaled so that $m=1$, $\hat{p}$ is the momentum operator, $\omega$ is the angular frequency of a linear oscillator, $\beta$ is the nonlinear parameter, and $g$ is the magnitude of an external periodic drive. This system has been widely studied in relation to chaotic dynamics in open quantum systems and the quantum-classical transition \cite{Sch1995,Bru1996,Ral2017a,Bha2003,Eve2005,Eas2016,Pok2016}. The full quantum (Q), as well as the corresponding classical (C) model, are of the form given in Eqs.~(\ref{eq:dynamics}) and (\ref{eq:results}), with additional dissipator terms to describe the interactions with gas particles, acting as a thermal environment (see Supplementary S2 for more details).

For the case considered here, one would like to find the conditions where one can best discriminate between the two dynamical models, and thus plan the experimental implementation accordingly. A number of different conditions were examined, for single well (linear and nonlinear) and double well potentials. The optimum condition was found to be a double-well potential with $\omega=1$, $\beta=0.5$ and $g=0$ (see Supplementary S3 for more details).
\begin{figure}
\includegraphics[width=0.9\columnwidth]{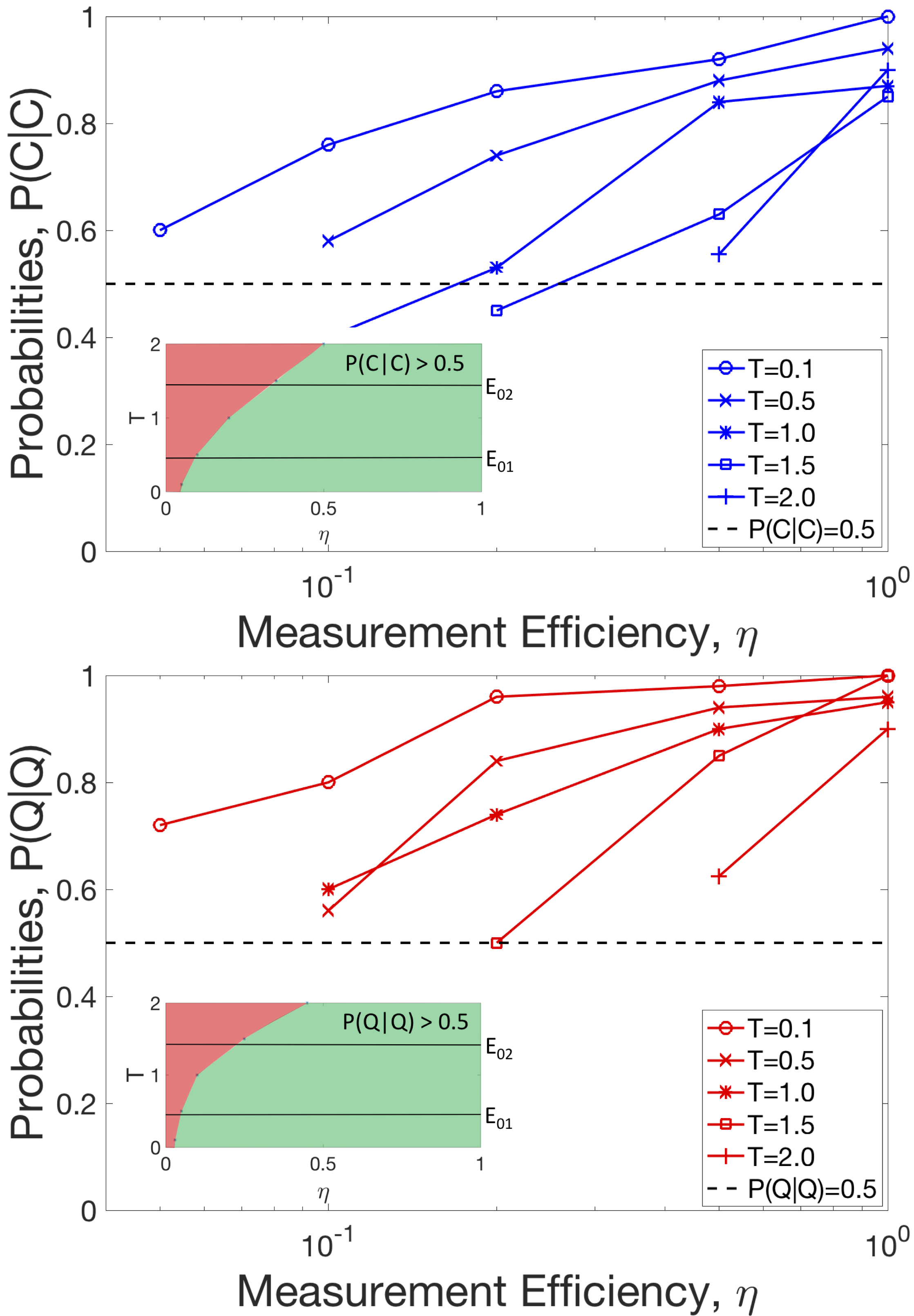}\caption{Numerically calculated probabilities for the identification of the correct dynamical evolution of optomechanical example with trapped nanoparticle in a double well potential: $Q$ is calculated using a quantum SME, and $C$ is calculated using the equivalent classical SDE. Probabilities $p({\rm C}|{\rm C})$ and $p({\rm Q}|{\rm Q})$ are shown, for different temperatures $T$ as a function of the measurement efficiency $\eta$. Inserts show approximate regions where the models are distinguishable (green (light gray) shaded regions), as functions of temperature and measurement efficiency.}
\label{fig:temperaturevariations}
\end{figure}

In general, the probabilities for correctly identifying a quantum system, $P(Q|Q)$, are slightly higher than for the classical system, $P(C|C)$. At low temperatures $k_{B}T<\Delta E_{01}$ the distinguishability is excellent, approaching 100\% even for measurement efficiencies $\eta\simeq0.2$, where $\Delta E_{01}$ is the energy separation between the ground state and the first excited state. This contrasts with a linear trap, where the probability of correctly distinguishing dynamical models was found to be limited to around 80\%, even for very low temperatures and ideal measurements $\eta=1.0$. Here, with two wells, both dynamical models show good distinguishability between quantum and classical behavior for temperatures $T\simeq0.5$ ($k_{B}T\simeq\Delta E_{01}$) and measurement efficiencies $\eta>0.2$, with some ability to distinguish between the two models for temperatures where the thermal energy is well above the first energy level separation and around the second transition, $k_{B}T\simeq1.5\hbar\omega\sim4\Delta E_{01}$, as long as $\eta>0.5$ (see Fig.~\ref{fig:temperaturevariations}).

Typical trapping frequencies in experiments are around $100$kHz and masses of the nanoparticles are a few $\times10^{-19}$kg \cite{Mud2016,Mud2017}. In this case, $\hbar\omega$ corresponds to a temperature of $0.77\mu$K, and $T=1.5\simeq1.16\mu$K, with the two wells separated by $0.2$nm, smaller than the radius of the sphere. However, double-well optical traps can be generated using fabricated structures within a few nanometers spacing between the two wells, below the diffraction-limit \cite{Tan2013,Kot2013}. Similarly, a few nanometers of spacing in ion trapping using optical lattices is demonstrated in \cite{Wang2017}. Alternatively, a double-well can be generated by focusing two laser beams of different wavelengths~\cite{Ron2017}. A dielectric particle will thus evolve in an effective potential of these two partially overlapping potentials. As highlighted in \cite{Tan2013,Vov2017} trapped particles can have resolutions well below $\sim1$pm \cite{Mud2016,Mud2017}. Therefore, such double-well traps are realisable within the current experiments. Experiments with levitated nanoparticles have reported temperatures around 450$\mu$K \cite{Jai2016}, well above the regime required, but experimental techniques are improving rapidly and temperatures equivalent to $\bar{n}\sim10-20$ are anticipated in the near future. Measurement efficiencies are more difficult to estimate from previous work since the values are not critical to the results presented and are not normally provided. However, for other systems, such as superconducting circuits \cite{Mur2013,Web2014,Six2015,Cam2016}, it is known that measurement efficiencies of at least $\eta\sim0.4$ are achievable \cite{Web2014}. 

\prlsec{Conclusions} This letter has discussed a general method to distinguish between dynamical models for quantum and classical systems. It provides an alternative to standard statistical tests based on state-reconstruction. We have re-phrased the problem of model selection in a form suitable to apply the well-known Neyman-Pearson decision rule, and quantified the reliability of the selection using the confusion matrix. Particularly noteworthy is the simplicity and generality of the proposed method: dynamical model selection is based on a generic time-trace data and it could be used to select between a wide variety of dynamical models. 

To illustrate the method of dynamical model selection, we have considered its application to levitated optomechanical systems, where non-classical features are yet to be experimentally demonstrated. We have proposed and optimized a novel experiment, where the nanoparticle is optically trapped in a double-well potential. Using dynamical model selection we have provided limits for two key experimental parameters (temperature and measurement efficiency) for quantum behaviour to be detected reliably. The successful experimental implementation, were it to confirm non-classical features, would improve on the most massive particle shown to exhibit quantum interference by several orders of magnitude~\cite{Eib2013}, and would thus be of great importance to fundamental physics.

\begin{acknowledgments}
JFR would like to thank H. M. Wiseman, P. Barker, and M. Flinders for useful and informative
discussions. MR, MT, AS and HU acknowledge funding by the Leverhulme Trust (RPG-2016-046)
and the Foundational Questions Institute (FQXi) and AS acknowledges
support by the Engineering and Physical Sciences Research Council
(EPSRC) under Centre for Doctoral Training grant EP/L015382/1.
\end{acknowledgments}

\bibliographystyle{apsrev}

\widetext

\section*{S1: General diffusive models}

In this supplementary we consider non-relativistic, Markovian, diffusive
models~\cite{Wis2010,Jac2014}. We start by discussing general classical models. Specifically, the conditional probability density $P_{c}$ evolves according to the
Kushner-Stratonovich equation \cite{Jac2014}:
\begin{eqnarray}
dP_{c} & = & -\sum_{k}^{r=1}\frac{\partial}{\partial q_{k}}\left(a_{k}P_{c}\right)dt\nonumber \\
 &  & +\frac{1}{2}\sum_{k}^{r=1}\sum_{k'=1}^{r}\frac{\partial}{\partial q_{k}\partial q_{k'}}\left(D_{kk'}P_{c}\right)dt\nonumber \\
 &  & +\sum_{k}^{r'}\sum_{k'=1}^{r'}\left[c-\mathbb{E}[c]\right]_{k}\left(BB^{\top}\right)_{kk'}\left[BdV+(c-\mathbb{E}[c])dt\right]_{k'}\label{eq:KS}
\end{eqnarray}
where $a_{k}\in\mathbb{R},$
$B_{kk'}\in\mathbb{R}$, $c_{k}\in\mathbb{R}$, $D_{kk'}\in\mathbb{R}$,
and $V$ is a vector of mutually independent $\mathbb{R}$-valued
Wiener processes. The first line corresponds to the Hamiltonian evolution,
the second line to the diffusion, i.e. in case the measurement perturbs
the system, and the third line to the update in the knowledge about
the system. In particular, the measured signal is given by (a $r'$-dimensional
vector):
\begin{equation}
dI_{\text{exp}}=cdt+BdV.
\end{equation}

We next describe general quantum models. Specifically, the conditional
statistical operator $\hat{\rho}_{c}$ evolves according to the Belavkin
equation \cite{Bel1999,Wis2010,Jac2014} :
\begin{eqnarray}
d\hat{\rho}_{c} & = & -i\left[\hat{H},\rho_{c}\right]dt+\mathcal{D}[\hat{c}]\hat{\rho}_{c}dt+\mathcal{H}[dU^{\dagger}\hat{c}]\hat{\rho}_{c},\label{eq:Belavkin}
\end{eqnarray}
where $\hat{c}$ is a $r'$-dimensional vector of operators. $U$ is
a $r'$-dimensional vector of correlated $\mathbb{C}$-valued Wiener
processes satisfying:
\begin{flalign}
dUdU^{\dagger} & =\eta dt,\\
dUdU^{\top} & =\varXi dt,
\end{flalign}
where $\eta$ is diagonal with $\eta_{kk}\in[0,1]$, and $\varXi$
is symmetric with $\mathbb{C}$-valued elements. Moreover, we have
the constraint that
\begin{equation}
\left[\begin{array}{cc}
\eta+\text{Re}(\varXi) & \text{Im}(\varXi)\\
\text{Im}(\varXi) & \eta+\text{Re}(\varXi)
\end{array}\right]
\end{equation}
is postive semi-definite. Note that the first, second, and third term
on the right hand-side of Eq.~(\ref{eq:Belavkin}) correspond to
the first, second, and third line of the right hand-side of Eq.~(\ref{eq:KS}),
respectively. The measurement signal is given by (a $r'$-dimensional
vector with $\mathbb{C}$-valued elements):
\begin{equation}
dI_{\text{exp}}=\text{Tr}\left[(\hat{c}^{\top}\eta+\hat{c}^{\dagger}\varXi)\hat{\rho}_{c}\right]dt+dU^{\top}.
\end{equation}

\section*{S2: Optomechanical system models }

For our purposes, the important factors are: (i) a levitated nanoparticle
is physically large (with a radius several hundred to a few thousand
times that of an atom); (ii) a nanoparticle has a high mass (six to
eight orders of magnitude larger than an atom); (iii) the trap can
be arranged to separate degrees of freedom in terms of frequency,
thereby simplifying the system to one translational degree of freedom;
and (iv) the particle is weakly coupled to a thermal environment and
to a laser field that can be used to provide a continuous measurement
of position. We will take parameters based on optomechanical spheres
described in \cite{Mud2016,Mud2017}, made from silica with radii
$\simeq25-100$nm and masses $m\simeq10^{-19}-10^{-18}$kg. These
are good candidates for study because they previously have been used
in experiments to generate thermal squeezed states \cite{Mud2016},
measurements have been used to reconstruct (classical) Wigner functions
\cite{Mud2017}, and they can realize the multiple-well potentials
\cite{Ron2017}, which we find maximizes the discrimination between
classical and quantum models.

A continuous quantum measurement process is usually modeled with a
Stochastic Master Equation (SME) \cite{Bel1999,Wis2010,Jac2014},
which can be written as 
\begin{eqnarray}
d\rho_{c} & = & -i\left[\hat{H},\rho_{c}\right]dt\nonumber \\
 &  & +\sum_{r=1}^{m'}\left\{ \hat{L}_{r}\rho_{c}\hat{L}_{r}^{\dagger}-\frac{1}{2}\left(\hat{L}_{r}^{\dagger}\hat{L}_{r}\rho_{c}+\rho_{c}\hat{L}_{r}^{\dagger}\hat{L}_{r}\right)\right\} dt\nonumber \\
 &  & +\sum_{r=1}^{m'}\sqrt{\eta_{r}}\left(\hat{L}_{r}\rho_{c}+\rho_{c}\hat{L}_{r}^{\dagger}-\mathrm{Tr}(\hat{L}_{r}\rho_{c}+\rho_{c}\hat{L}_{r}^{\dagger})\right)dW_{r},\label{sme1}
\end{eqnarray}
where $\rho_{c}$ is the density matrix for the state of the system
conditioned on the measurement record \textendash{} the state (possibly
mixed), which represents the current knowledge of the quantum state,
$\hat{H}$ is the Hamiltonian of the system, $dt$ is an infinitesimal
time increment, and the operators $\hat{L}_{r}$ represent the effect
of the environment and measurement. The measurement record for each
of the operators $\hat{L}_{r}$ during a time step $t\rightarrow t+dt$
is given by, $y(t+dt)-y(t)=dy_{r}(t)=\sqrt{\eta_{r}}\mathrm{Tr}(\hat{L}_{r}\rho_{c}+\rho_{c}\hat{L}_{r}^{\dagger})dt+dW_{r}$, where the recorded time trace in this interval is $dI_\text{exp}(t) =dy(t)/ \sqrt{2k}$.
$\eta_{r}$ is the measurement efficiency; the ratio of the signal
power due the measurement relative to the power of other extraneous
sources of noise, where $\eta_{r}=1$ is an ideal measurement and
$\eta_{r}=0$ is an unprobed environmental degree of freedom. Moreover,
we will assume that $dW_{r}$ are independent real Wiener processes,
i.e. $dW_{r}dW_{r'}=\delta_{rr'}dt$. Physically, this SME represents
a situation where the measurement environment decoheres sufficiently
rapidly that no correlations build up between the state of the quantum
system of interest and the environmental degrees of freedom (Markov
approximation).

For the case considered here, the SME is given by (\ref{sme1}) with
three environmental operators ($m'=3$): one measurement of the position
($q$) of the nanosphere within the trap, $\hat{L}_{1}=\sqrt{2k}\hat{q}$,
and two operators representing an unprobed thermal environment $\hat{L}_{2}=\sqrt{(\bar{n}+1)\Gamma}\hat{a}^{\dagger}$
and $\hat{L}_{3}=\sqrt{\bar{n}\Gamma}\hat{a}$ \cite{Spi1993}; where
$\hat{a}^{\dagger}$ and $\hat{a}$ are the usual harmonic oscillator
raising and lowering operators, $\Gamma$ is a decay rate ($\Gamma\ll\omega$),
$\bar{n}=(\exp(\hbar\omega/k_{B}T)-1)^{-1}$ is the average thermal
occupation number of a linear oscillator at temperature $T$, and
$k$ is the measurement strength for the continuous measurement interaction.
The measurement efficiencies are $\eta_{1}=\eta$, and $\eta_{2,3}=0$
(unprobed). The measurement record for $\hat{L}_{1}$ is $dy(t)=\sqrt{8\eta k}{\rm Tr}(\rho_{c}(t)\hat{q})dt+dW$. 

For the equivalent classical system, we take a Stochastic Differential
Equation (SDE) for the position $q$ and the momentum $p$ of a classical
particle, 
\begin{eqnarray}
dq & = & pdt\nonumber \\
dp & = & -\mu q^{3}dt+\omega^{2}qdt-\Gamma pdt+g\cos(t)\nonumber \\
 &  & +\sqrt{2k}dY+\sqrt{2\Gamma k_{B}T}dU\label{sde1}
\end{eqnarray}
where the measurement record is $dy_{c}(t)=\sqrt{8\eta k}qdt+dW$
and we have again set $\hbar=1$. Like $dW$, $dY$ and $dU$ are
also real Weiner increments, $dY^{2}=dU^{2}=dt$, but they are uncorrelated
so that $dWdU=dWdY=dYdU=0$, and there is no backaction from the measurement
on the state of the system in a classical measurement.

\section*{S3: Optomechanical system simulation}

The distinguishibility of the two models was found to be best in the
double well configuration. Specifically, when the two wells were well
separated in position and the barrier between the two wells was high
enough for the classical particle to remain in one well for a reasonable
period of time, before the environmental noise kicked it into the
other well. In addition, the barrier also had to be low enough to
prevent the quantum state localizing in one or other of the wells.
In practice, these conditions correspond to a symmetric double well
potential where the quantum ground state lies below the barrier height
but the first excited state is above the barrier. The classical system
is always localized, in the sense that it is a point particle, but
the pdf represented by the particles needs to be largely localized
to one of the wells by the measurements. By contrast, a quantum state
can only be localized to one of the wells if two of the low lying
energy levels are below the barrier. If the barrier is sufficiently
high, the lowest two energy states are formed from the symmetric and
anti-symmetric superposition of localized well states, and a localized
well state can be generated by combining these two energy levels \cite{Eve2004}.
If the first excited state lies above the barrier, then a superposition
of this with the ground state will not be localized in one well. For
the Duffing Hamiltonian (\ref{DuffHam}), these conditions are met
if we take $\omega=1$, $\beta=0.5$ and $g=0$, where we have set
$\Gamma=0.05$, $k=0.025$, $N=500$. 

The quantum model uses Rouchon's integration method \cite{Ami2011,Rou2015}
with non-commutative noise sources and a moving basis \cite{Sch1995,Bru1996,Ral2017a}
with 60-100 linear oscillator states. The models are integrated over
100 cycles of the linear oscillator with 500-2000 time steps per oscillator
cycle, and the probabilities are calculated bu averaging over 100 runs of each model. 
The barrier height in this example is $\Delta E_{b}=0.5\hbar\omega$,
the two wells are separated in position by $\Delta q=3\sqrt{\hbar/m\omega}$,
the lowest two energy levels are separated by $\Delta E_{01}=0.396\hbar\omega$,
and the next excited states are separated by $\Delta E_{12}=0.941\hbar\omega$
and $\Delta E_{23}=1.061\hbar\omega$.

The classical model requires the evolution of the probability density
function (pdf) to be calculated, which is computationally expensive.
We use an alternative approach here to solve the approximate problem
using a sequential Monte Carlo method \cite{Gor1993,Dou2001,Aru2002}
known as a particle filter. The particle filter uses the fact that
the evolution of the pdf can be approximated by the evolution of a
finite number of candidate solutions or `particles', each of which
has a weight associated with it, where the weight evolves in such
a way that a quantity averaged over all weighted particles approximates
the expectation value for the quantity over the pdf. In this case,
we take $N$ particles, initialized with equal weight $w_{0}^{(i)}=1/N$.
Each particle has a position $q^{(i)}$ and a momentum $p^{(i)}$,
initially selected from the same thermal distribution as that given
by the thermal state for the quantum model. The particles then evolve
according to the SDE (\ref{sde1}) with independent noise sources.
The weights are updated using 
\begin{equation}
\tilde{w}_{t}^{(i)}=\frac{\exp\left(-\frac{(\Delta\tilde{y}_{t}-\sqrt{8\eta k}q^{(i)}\Delta t)^{2}}{2\Delta t}\right)}{\sqrt{2\pi\Delta t}}w_{t-\Delta t}^{(i)}
\end{equation}
where the $\tilde{w}^{(i)}$'s are unnormalized weights after updating,
and the probability for the classical model is approximated by $p(\Delta\tilde{y}_{t}|{\rm C},\Delta\tilde{y}_{0:t'-\Delta t})\propto\sum_{i=1}^{N}\tilde{w}_{t}^{(i)}$.
As the system evolves, the values of some of the weights fall to near
zero. The particles and the candidate solutions that they represent
are then resampled using the current weight distribution as described
in \cite{Gor2002}. This evolution with periodic resampling allows
the particle filter to be efficient whilst still retaining a diverse
selection of candidate solutions. This makes the particle filter an
ideal method for the estimation of a nonlinear dynamical process and
it is the reason for considering it in a model selection context.
In addition, the particle filter and other sequential Monte Carlo
methods can be augmented to include the simultaneous estimation of
system parameters \cite{Gre2017} and they can be applied to quantum
systems described by SMEs with uncertain parameters \cite{Ral2017b}.

It should be noted that the ability to distinguish the models is dependent
on the total time over which the measurement record is collected and
the models integrated. Extending the integration time will improve
the results, but the trap potential and the measurement interaction
would need to be stable over the integration time, providing a trade-off
between distinguishability and difficulties in collecting the measurement
data.
\end{document}